\documentclass[aps,amsmath,amssymb,prb,reprint,superscriptaddress]{revtex4-1}
\usepackage{verbatim}
\usepackage[colorlinks=true,linkcolor=blue,citecolor=blue,urlcolor=blue]{hyperref}
\usepackage{amsmath}
\usepackage{amssymb} 
\usepackage[pdftex]{graphicx}
\begin{document}

\title{Direct measurement of hyperfine shifts and radiofrequency manipulation of the nuclear spins in individual CdTe/ZnTe quantum dots}

\author{G. Ragunathan}\email{gragunathan1@sheffield.ac.uk}
\affiliation{Department of Physics and Astronomy, University of
Sheffield, Sheffield, S3 7RH, UK}
\author{J. Kobak}\email{Jakub.Kobak@fuw.edu.pl}
\affiliation{Department of Physics and Astronomy, University of
Sheffield, Sheffield, S3 7RH, UK} \affiliation{Institute of
Experimental Physics, Faculty of Physics,  University of Warsaw,
ul. Pasteura 5, 02-093 Warsaw, Poland}
\author{G. Gillard}
\affiliation{Department of Physics and Astronomy, University of
Sheffield, Sheffield, S3 7RH, UK}
\author{W. Pacuski}
\affiliation{Institute of Experimental Physics, Faculty of
Physics,  University of Warsaw, ul. Pasteura 5, 02-093 Warsaw,
Poland}
\author{K. Sobczak}
\affiliation{Biological and Chemical Research Centre, Faculty of
Chemistry, University of Warsaw, ul. Zwirki i Wigury 101, 02-089
Warsaw, Poland}
\author{J. Borysiuk}
\affiliation{Institute of Experimental Physics, Faculty of
Physics,  University of Warsaw, ul. Pasteura 5, 02-093 Warsaw,
Poland}
\author{M. S. Skolnick}
\affiliation{Department of Physics and Astronomy, University of
Sheffield, Sheffield, S3 7RH, UK}
\author{E. A. Chekhovich}\email{e.chekhovich@sheffield.ac.uk}
\affiliation{Department of Physics and Astronomy, University of
Sheffield, Sheffield, S3 7RH, UK}

\date{\today}

\begin{abstract}
We achieve direct detection of electron hyperfine shifts in
individual CdTe/ZnTe quantum dots. For the previously inaccessible
regime of strong magnetic fields $B_z\gtrsim0.1$~T, we demonstrate
robust polarization of a few-hundred-particle nuclear spin bath,
with optical initialization time of $\sim$~1~ms and polarization
lifetime exceeding $\sim$~1~s. Nuclear magnetic resonance
spectroscopy of individual dots reveals strong electron-nuclear
interactions characterized by the Knight fields
$|B_e|\gtrsim50$~mT, an order of magnitude stronger than in III-V
semiconductor quantum dots. Our studies confirm II-VI
semiconductor quantum dots as a promising platform for hybrid
electron-nuclear spin quantum registers, combining the excellent
optical properties comparable to III-V dots, and the dilute
nuclear spin environment similar to group-IV semiconductors.
\end{abstract}

\maketitle

\newcommand{\RedText}[1]{\textcolor{red}{#1}}
\newcommand{\BlueText}[1]{\textcolor{blue}{#1}}
\newcommand{\GreenText}[1]{\textcolor{green}{#1}}

\begin{figure}[h]
\includegraphics[width=0.99\columnwidth]{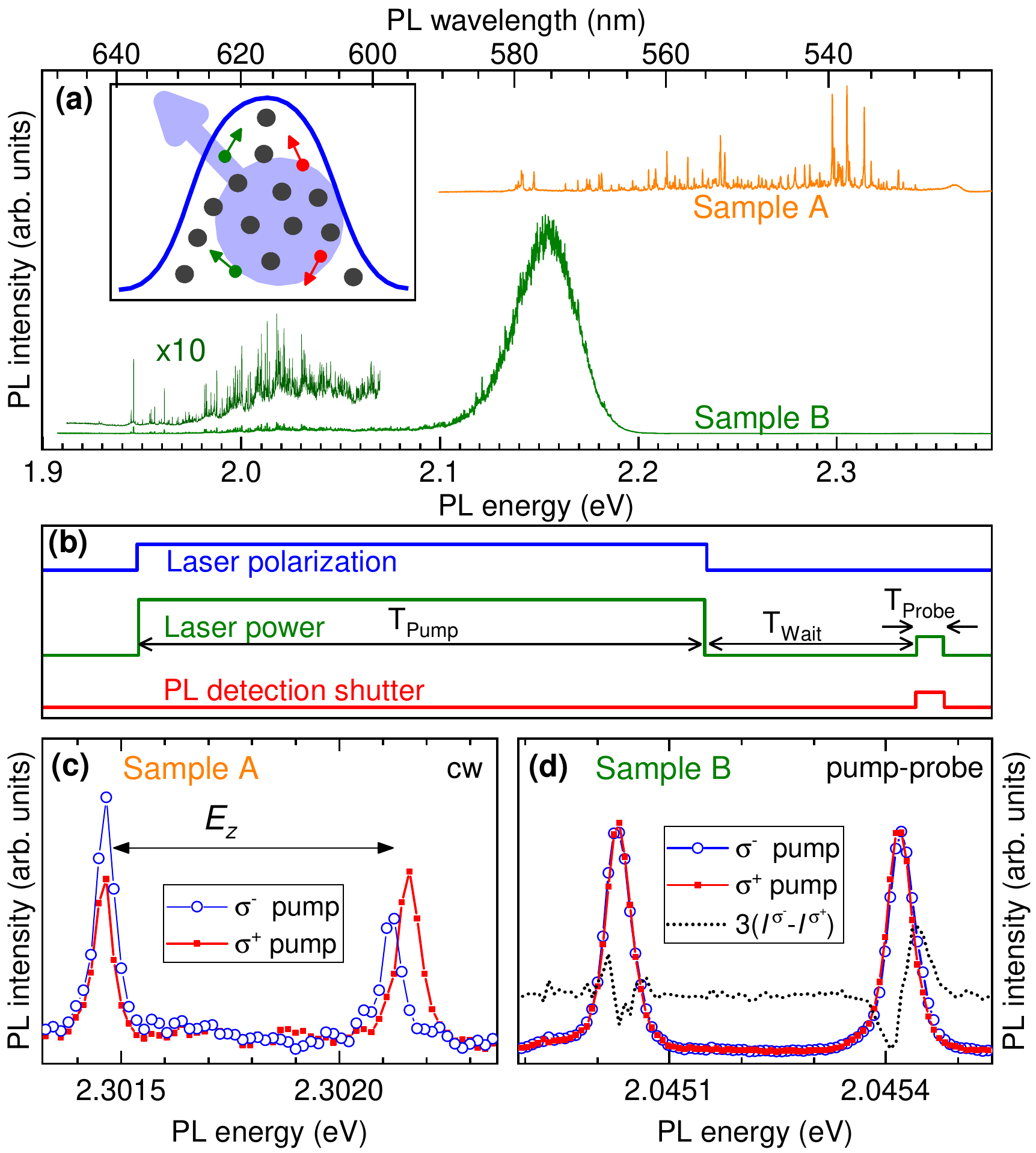}
\caption{(a) Broad range photoluminescence (PL) spectra from
CdTe/ZnTe samples A (top) and B (bottom) under 405~nm laser
excitation. PL spectrum of sample A consists of QD emission at
2.13 -- 2.34~eV and a weak ZnTe barrier peak at 2.36~eV. Sample B
spectrum includes QDs at 1.94 -- 2.09~eV and CdTe quantum well
emission centered at 2.155~eV. Inset shows schematic of the
quantum dot electron spin (large arrowhead) interacting with only
a few spin-1/2 nuclei (small arrowheads) randomly distributed
among spin-0 nuclei (small circles). (b) Timing of the pump-probe
measurement cycle, where laser polarization, power, and PL
detection are controlled separately to implement pump and probe
pulses with different characteristics. (c) Continuous wave (cw) PL
spectra of a single CdTe/ZnTe neutral QD from sample A under
$\sigma^{-}$ (circles) and $\sigma^{+}$ (squares) excitation at
$B_{z}=$~4~T. A visible change in splitting $E_{Z}^{\sigma^{-}} -
E_{Z}^{\sigma^{+}}\approx$~$-44$~$\mu$eV is detected but is not
related to nuclear spin effects. (d) Pump-probe PL spectra of a
single CdTe/ZnTe neutral QD from sample B measured at
$B_{z}=$~2.5~T with linearly polarized probe and $\sigma^{-}$
(circles) or $\sigma^{+}$ (squares) pump using
$T_\textrm{Pump}=40$~ms, $T_\textrm{Wait}=4.5$~ms,
$T_\textrm{Probe}=1$~ms. The change in Zeeman splitting
$E_{Z}^{\sigma^{-}} - E_{Z}^{\sigma^{+}}\approx$~4~$\mu$eV is
smaller than the linewidth (36~$\mu$eV), but is revealed in the
difference spectrum (dotted line, $\times$3 scaled for clarity)
and is ascribed to dynamic nuclear
polarization.}\label{fig:PLSpec}
\end{figure}

\begin{figure}
\includegraphics[width=\columnwidth]{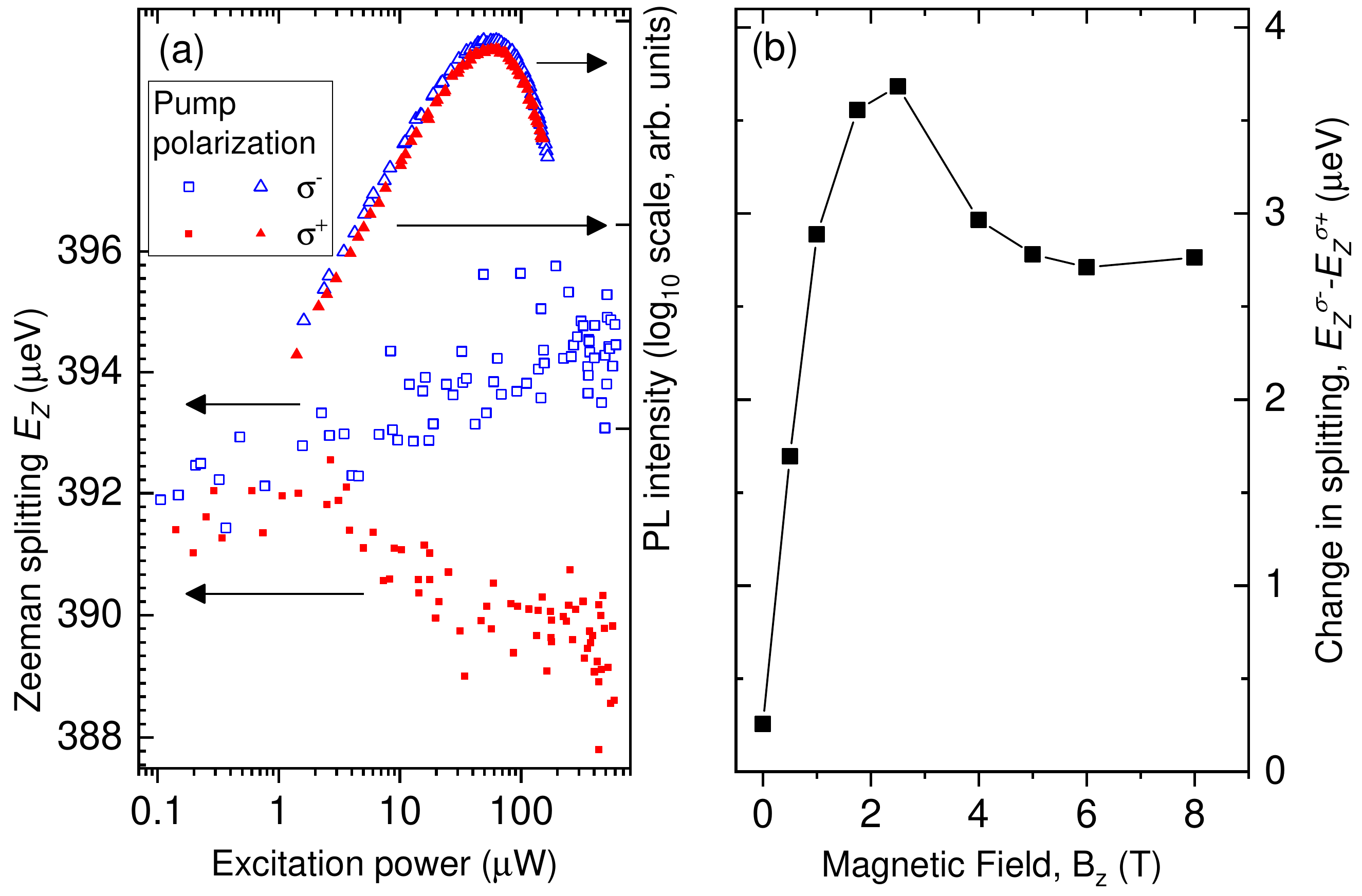}
\caption{(a) Total PL intensity (sum for two Zeeman components) of
a QD from sample B measured in cw (triangles, right scale), and
Zeeman splitting $E_{Z}$ measured in pump-probe (squares, left
scale) as a function of power under $\sigma^{-}$ (open symbols)
and $\sigma^{+}$ (solid symbols) 561~nm laser excitation at
$B_{z}$ = 2.5~T. (b) Magnetic field dependence of the change in
Zeeman splitting $E_{Z}^{\sigma^{-}} - E_{Z}^{\sigma^{+}}$
measured on the same quantum dot in pump-probe experiment with
pump duration $T_\textrm{Pump}=$~40~ms and power
160~$\mu$W.}\label{fig:OHSproperties}
\end{figure}

\newcommand{\FigBroadPL}{\ref{fig:PLSpec}(a)\:}
\newcommand{\FigTDiagSpec}{\ref{fig:PLSpec}(b)\:}
\newcommand{\FigDNPA}{\ref{fig:PLSpec}(c)\:}
\newcommand{\FigDNPB}{\ref{fig:PLSpec}(d)\:}
\newcommand{\FigOHSPDep}{\ref{fig:OHSproperties}(a)\:}
\newcommand{\FigOHSBDep}{\ref{fig:OHSproperties}(b)\:}
\newcommand{\FigDNPRise}{\ref{fig:nucDynamics}(a)\:}
\newcommand{\FigDNPDec}{\ref{fig:nucDynamics}(b)\:}

The proposed designs for solid-state quantum information
processing devices require two essential components: the quantum
nodes for storing and processing information, and the quantum
channels connecting the nodes \citep{Kimble2008}. Various material
systems using single spins as qubit nodes and single photons as
channels have been considered. III-V semiconductor quantum dots
(QDs) are of particular interest, since they benefit from mature
epitaxial technologies and exceptional single-photon properties
\citep{SomachiNPhot2016,PhysRevLett.116.020401,PhysRevLett.119.010503}.
However, the electron spin qubits \citep{Loss1998} suffer fast
decoherence due to the interaction with a dense nuclear spin
environment \citep{Urbaszek2013,NatMatReview2013}. By contrast,
group IV semiconductors, such as silicon and diamond, where most
nuclei are spin-free ($I = 0$) offer defect-spin qubits with
record coherence \citep{Childress2006,Pla2012}, while their
optical properties and fabrication technology are inherently
limited. The advantages of the two approaches can be combined if
optically active quantum dots can be grown of materials with
spin-free nuclei. The II-VI semiconductors are a natural choice
for this since most nuclei are spin-free and the direct-bandgap
character offers a good interface between electron spin and
photons.

The research of the past two decades have lead to an in-depth
understanding and development of advanced techniques for probing
and manipulation of the nanoscale nuclear spin ensembles in III-V
QDs \citep{Urbaszek2013,NatMatReview2013}. By contrast, current
understanding of the nuclear spin phenomena in II-VI dots is
scarce, due to the challenges arising from the small nuclear spin
magnetisation in a dilute spin bath. Previous studies
\citep{Akimov2006,PhysRevLett.99.036604,Kim2010,LeGall2012} relied
on indirect detection of the nuclear spin effects via probing of
the electron spin dynamics. Consequently, these experiments were
restricted to low magnetic fields ($B\lesssim0.1$~T), leaving
beyond reach the most interesting regime where nonsecular
electron-nuclear spin interactions are suppressed by magnetic
field giving access to long-lived electron and nuclear spin
states, required for qubit applications.

Here we achieve direct probing of the nuclear spin state by
measuring Overhauser spectral shifts in individual CdTe/ZnTe
quantum dots, which enables studies in a wide range of external
fields. A cascade relaxation process involving quantum well states
is identified as a source of efficient dynamic nuclear
polarization (DNP) in magnetic fields up to 8~T. The DNP can be
induced within $\sim$~1~ms and persists in the dark over
$\sim1$~s, three orders of magnitude longer than observed
previously in II-VI QDs at low fields. The direct detection of
spectral shifts employed here, reveals an additional effect which
mimics DNP, but is characterized by submicrosecond timescales
pointing to electron spin interactions as a source. While in
previous studies nuclear species could not be addressed
individually, here we measure cadmium and tellurium nuclear
magnetic resonance signals in individual CdTe dots and observe
strong electron-nuclear interaction characterized by the Knight
fields exceeding 50~mT. Our results suggest CdTe/ZnTe quantum dots
as a promising system with a potential of implementing a hybrid
quantum spin register architecture \citep{Dutt2007} based on one
electron coupled to few individually addressable nuclei, and with
high optical efficiency unachievable in group-IV semiconductors.


We study two CdTe/ZnTe samples grown by molecular beam epitaxy. In
sample A, low-density QDs were formed using the amorphous Te
technique \citep{Tinjod2003,JKobakJCG2013}, whereas in sample B,
amorphous Te deposition was avoided, resulting in a higher QD
density and preservation of the CdTe wetting layer quantum well
(for further details see Supplemental Material Note 1).
Micro-photoluminescence ($\mu$-PL) experiments are conducted at a
temperature of 4.2~K with an external magnetic field $B_{z}$
applied along the sample growth axis (Faraday geometry). PL of
individual QDs is excited non resonantly using a solid-state laser
emitting at 532~nm or 561~nm, and the emission of a neutral
exciton (X$^{0}$) state is collected and dispersed by a 1~m double
spectrometer, followed by a pair of achromatic doublets, which
transfers the spectral image onto the charge coupled device (CCD)
detector with a linear magnification of 3.75. Using Gaussian
fitting it is possible to detect the change in splitting of
spectral peaks with an accuracy of $\sim$~0.5~$\mu$eV. In order to
implement pump-probe measurements, the polarization of the laser
is modulated with an electro-optical modulator (Pockels cell,
response time $\approx$~0.5~$\mu$s), while analogue modulation of
the diode pump current is used to modulate the power of the laser
(response time $<$~1~$\mu$s) from the nominal power to zero. The
PL signal is modulated with an acousto-optic modulator (response
time $\approx$~1.1~$\mu$s, on/off ratio $>$~1000) or a liquid
crystal cell (response time $\approx$~4.5~ms, on/off ratio
$>$~5000).


Fig.~\FigBroadPL shows broad-range PL spectra of samples A and B.
Both structures exhibit sharp spectral lines characteristic of QD
emission with full width at half maximum as low as
$\sim$~30~$\mu$eV in sample A and $\sim$~20~$\mu$eV in sample B.
For sample B, a strong, broad PL peak arising from QW emission is
observed at $\sim$~2.15~eV, and the number of sharp peaks is an
order of magnitude higher than in sample A, confirming the higher
dot density in sample B. In most studied individual QDs, PL is
dominated by recombination of a bright neutral exciton, recognized
through its fine structure splitting. In an external magnetic
field $B_z$, the bright exciton localized in a QD becomes a
doublet of states with electron spin parallel or antiparallel to
$B_z$ and with Zeeman energy splitting $E_{Z}$. Nuclear spins
polarized along the $z$ axis act on the electron via the hyperfine
interaction and shift the exciton energies in opposite directions,
which can be observed in the PL spectrum as a change in $E_{Z}$
\citep{Urbaszek2013,NatMatReview2013}. In III-V QDs excitation
with $\sigma^+$ or $\sigma^-$ circularly polarized light results
in repeated injection of spin polarized electrons into the dot,
leading to DNP of the dot nuclei via hyperfine interaction. The
resulting variation in the exciton Zeeman splitting can be as
large as $|E_{Z}^{\sigma^{+}} - E_{Z}^{\sigma^{-}}|>$~200~$\mu$eV,
significantly exceeding the PL linewidths, and a pronounced effect
is observed \citep{Gammon2001,Urbaszek2013,NatMatReview2013}.

Using the same approach, we measure PL spectra under $\sigma^+$ or
$\sigma^-$ excitation at 532~nm in CdTe/ZnTe QDs. A typical result
for a single QD in sample A is presented in Fig. \FigDNPA where a
noticeable ($\sim$~44~$\mu$eV) change in $E_{Z}$ is detected.
Although this result is reminiscent of how DNP manifests in III-V
dots, in some CdTe/ZnTe dots variation of $E_{Z}$ exceeds that
expected for $\pm$~100$\%$ nuclear spin polarization. To verify
this effect, additional pump-probe measurements were conducted
with pulse timing shown in Fig.~\FigTDiagSpec. In this experiment,
the dot is periodically excited with a pump pulse with a variable
degree of circular polarization (e.g. $\sigma^+$, $\sigma^-$ or
linear), while the PL is detected only during a subsequent short
probe pulse with a fixed linear polarization. The key feature of
this experiment is that the observed splitting $E_{Z}$ is
sensitive only to those effects of the pump that persist over a
sufficiently long time $T_\textrm{Wait}$, as should be the case
for nuclear spin polarization whose lifetime is orders of
magnitude longer than the exciton radiative recombination time of
a few hundred picoseconds.

While nearly 15\% out of $\sim$~90 individual QDs examined in
sample A showed a pronounced change in $E_{Z}$ in continuous wave
(cw) PL [see Fig.~\FigDNPA], surprisingly, no measurable change in
$E_{Z}$ could be detected in pump-probe experiments with either
532~nm or 561~nm excitation. Moreover, additional dynamics
measurements have shown that the change in $E_{Z}$ induced by a
circularly polarized pump persists for less than
$T_\textrm{Wait}<0.5$~$\mu$s (limited by modulators resolution),
which is too short to be ascribed to nuclear spin polarization
dynamics. The exact origin of the changes in $E_{Z}$ in cw PL is
not yet clear and requires further investigation. The effect is
absent for 561~nm excitation but is observed under 532~nm
excitation, and only for the dots with small detuning from the
laser (QD ground state emission between 537~nm and 539~nm). Based
on the sub-microsecond timescales it is likely to be related to
electron or hole spin effects, for example dipolar and/or exchange
interaction with spin polarized charges in nearby quantum dots or
defects.

Having established the absence of DNP in sample A, we have
conducted pump-probe experiments on QDs in sample B using 561~nm
pump laser, with a typical result presented in Fig.~\FigDNPB. The
change in Zeeman splitting $E_{Z}^{\sigma^{-}} -
E_{Z}^{\sigma^{+}}\approx$~4~$\mu$eV is smaller than the PL
linewidths, but is detected reliably from Gaussian lineshape
fitting. Similar results were obtained from the measurements on
$\sim$~20 individual QDs from sample B. The systematic nature and
the sign (see Supplemental Note 3) of the shift observed in
pump-probe measurements suggests DNP as its origin. Further
investigation is presented in Fig.~\FigOHSPDep where power
dependent measurements are shown: at low power, $E_{Z}$ (squares)
does not depend on polarization of the pump, but at higher power,
a clear increase (decrease) in $E_{Z}$ is observed under
$\sigma^-$ ($\sigma^+$) pumping, saturating above
$\sim$~50~$\mu$W, which corresponds to the saturation power of the
bright neutral exciton PL intensity (triangles). Such saturation
is observed in all studied dots in sample B, which is different
from the III-V QDs (InGaAs/GaAs, GaAs/AlGaAs, InP/GaInP) where DNP
under non-resonant optical excitation is often found to be most
efficient at optical powers significantly exceeding the saturation
of the ground state \citep{DNPInP,DNPInGaAs,ChekhovichPRB2016},
implying the role of multi-exciton and excited QD states.

Fig.~\FigOHSBDep shows the dependence of the DNP measured as the
difference $E_{Z}^{\sigma^{-}} - E_{Z}^{\sigma^{+}}$ in a QD in
sample B at different magnetic fields $B_z$. DNP is nearly absent
at zero field but becomes more efficient with applied field,
reaching a maximum at $B_z\approx2.5$~T. Such an increase with
$B_z$ can be explained as follows: At $B_z=0$~T, bright exciton
states have no electron spin polarization due to the fine
structure splitting ($\delta_\textrm{b}\approx$~115~$\mu$eV for
this dot) and thus can not interact with nuclear spins. With
applied magnetic field, exciton Zeeman splitting increases
($\approx$~150~$\mu$eV/T for this dot, see Supplemental Note 2)
restoring electron spin polarization of the exciton states and
re-enabling interaction with the nuclei. At large magnetic fields
significant DNP is observed up to $B_z=8$~T, the partial reduction
of DNP above 2.5~T is similar to that observed in III-V QDs
\citep{Chekhovich2010} and is likely due to the mismatch in the
electron and nuclear spin energy splitting, which increases with
magnetic field, slowing down DNP.

Experimental observations presented above allow to make general
conclusions about the mechanism of DNP. Saturation of the nuclear
spin polarization level with saturation of the QD ground state PL,
as well suppression of DNP at low magnetic fields where fine
structure splitting dominates point to the key role of the neutral
exciton spin states. On the other hand, DNP is found to be
efficient only in structure B, where quantum well states are
present in addition to QD states. This is despite the fact that
circularly polarized optical excitation produces similar exciton
spin polarization degrees in both structures: exciton spin
polarization is evidenced in Fig.~\FigDNPA where $\sigma^+$
($\sigma^-$) excitation enhances PL intensity $I_\textrm{h}$
($I_\textrm{l}$) of the high (low) energy exciton state, and the
difference in PL polarization degrees
$\rho=(I_\textrm{h}-I_\textrm{l})/(I_\textrm{h}+I_\textrm{l})$
under $\sigma^+$ and $\sigma^-$ excitation is
$\Delta\rho\approx(+0.11)-(-0.35)=0.45$. We find similar
$\Delta\rho$ values for QDs in both samples while DNP is observed
only in sample B, suggesting that DNP takes place not during
ground state radiative lifetime or recombination, but rather
during the initial relaxation and formation of the QD ground state
exciton. This interpretation is supported by the presence of the
intermediate quantum well states in sample B, which appear to be a
critical factor for efficient spin exchange between the electrons
and nuclei.

\begin{figure}
\includegraphics[width=\columnwidth]{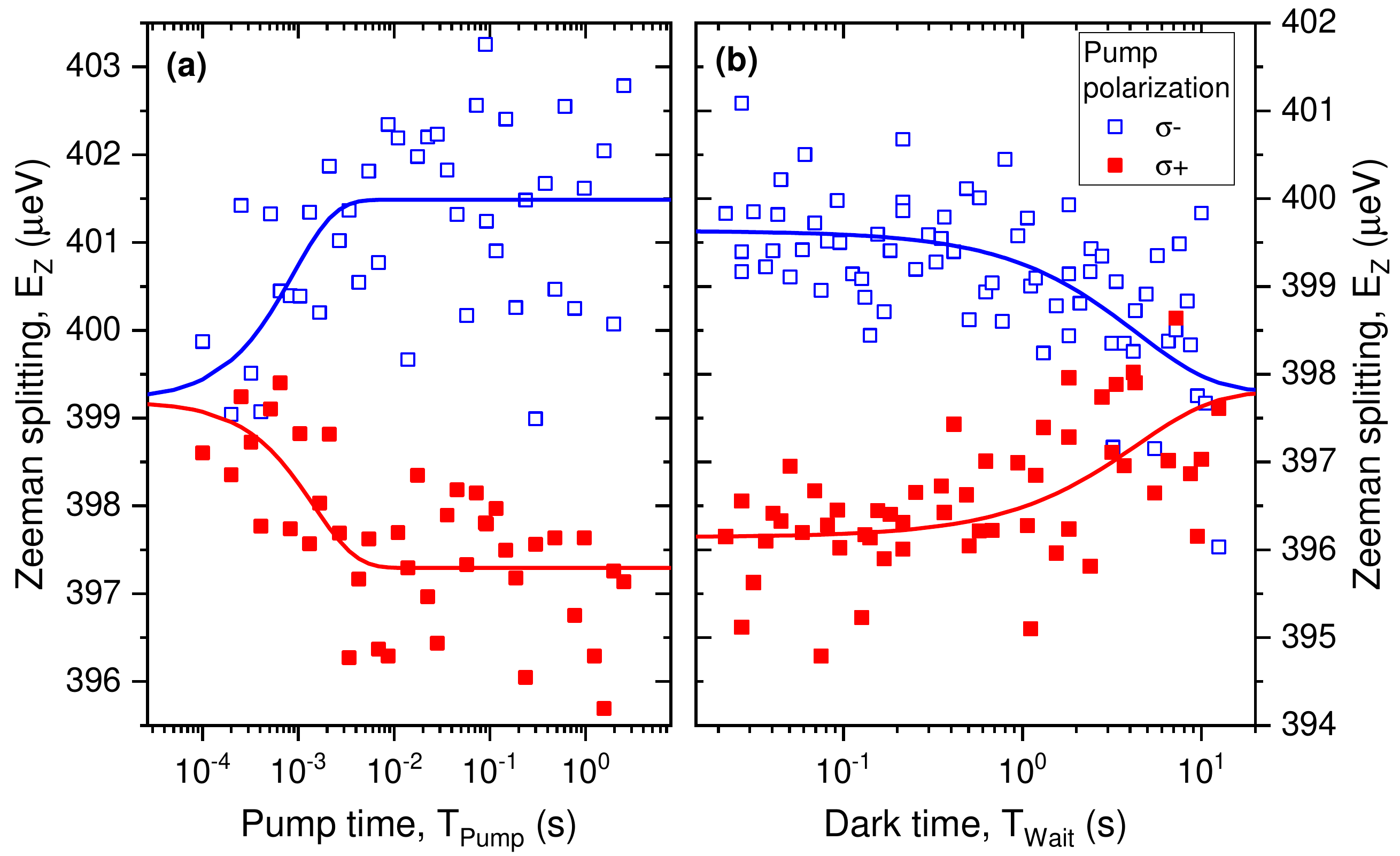}
\caption{(a) Buildup dynamics of the optically induced nuclear
spin polarization in a single QD at $B_{z}=$~2.5~T, measured with
a pump-probe cycle shown in Fig.~\FigTDiagSpec, but with the
addition of an erase laser pulse (duration
$T_\textrm{Erase}=$~50~ms, power 160 $\mu$W, linearly polarized)
preceding each pump pulse. Probe pulse power and duration are
20~$\mu$W and $T_\textrm{Probe}=1$~ms. Open (solid) symbols show
experiment with $\sigma^{-}$ ($\sigma^{+}$) polarized pump, lines
show exponential fitting yielding 95$\%$ confidence intervals for
the buildup time: $\tau_\textrm{Buildup}=0.9^{+1.3}_{-0.5}$~ms
($1.5^{+3.0}_{-1.0}$~ms). (b) Decay of the nuclear spin
polarization measured using the cycle of Fig.~\FigTDiagSpec with
variable dark time $T_\textrm{Wait}$ and fixed
$T_\textrm{Pump}=40$~ms, $T_\textrm{Probe}=1$~ms. The nuclear spin
lifetime derived from the exponential fit (lines) is
$\tau_\textrm{Decay} = 4.3^{+2.5}_{-1.6}$ s (95$\%$
confidence).}\label{fig:nucDynamics}
\end{figure}

Focusing on sample B we now turn to investigation of the nuclear
spin dynamics in individual CdTe/ZnTe QDs. Open (solid) symbols in
Fig. \FigDNPRise show the buildup dynamics of the DNP under
$\sigma^{-}$ ($\sigma^{+}$) pumping measured at $B_z=2.5$~T, where
exponential fitting (lines) reveals characteristic build up time
of $\tau_\textrm{Buildup}\sim$~1~ms. Similar
$\tau_\textrm{Buildup}$ were observed on different individual QDs
at $B_z=2.5$~T. These $\tau_\textrm{Buildup}$ values are a factor
of $\sim1000$ smaller than $\tau_\textrm{Buildup}\sim$~0.5 -- 3~s
found in III-V QDs at high magnetic fields \citep{Chekhovich2010,
Nikolaenko2009,Belhadj2008,ChekhovichPRB2016}. One of the
contributions to shorter $\tau_\textrm{Buildup}$ is the smaller
abundance of magnetic isotopes ($\sim25\%$ for Cd and $\sim8\%$
for Te, compared to $100\%$ for group III and V nuclei) and
smaller spin number $I=1/2$ (as opposed to $I=3/2$ for Ga and As,
and $I=9/2$ for In), which requires fewer electron-nuclear
flip-flops to approach equilibrium nuclear spin polarization in
CdTe QDs. However, based on lower $I$ and abundance alone, one
would expect a factor of $\sim30$ shorter $\tau_\textrm{Buildup}$
for CdTe. The remaining difference is due to a smaller QD volume,
typically containing $\sim$~5$\times$10$^{3}$ atoms (based on
transmission electron microscopy, see Supplementary Note 1), as
opposed to 10$^{4}$-10$^{5}$ atoms in III-V QDs. The buildup time
$\tau_\textrm{Buildup}\sim1$~ms observed here at $B_{z}$ = 2.5 T
is an order of magnitude longer than the $<$ 100 $\mu$s time found
previously in CdTe/ZnTe \citep{LeGall2012} and CdSe/ZnSe
\citep{Akimov2006} dots at low magnetic fields, which is well
explained by the reduction of the electron-nuclear spin flip-flop
probability due to the increasing mismatch in the Zeeman energies.

The measurement of the nuclear spin polarization decay in the
dark, following excitation with a $\sigma^{-}$ ($\sigma^{+}$)
polarized pump is show in Fig. \FigDNPDec by the the open (solid)
symbols. At long waiting times nuclear polarization is seen to
decay almost completely, with characteristic time
$\tau_\textrm{Decay}\sim$~4~s deduced from exponential fitting
(lines). Similar $\tau_\textrm{Decay}$ were observed in several
individual CdTe quantum dots in sample B and are significantly
longer than submilisecond $\tau_\textrm{Decay}$ reported for
charged CdSe QDs at low magnetic fields
\citep{PhysRevLett.99.036604}, but are noticeably shorter than
$\tau_\textrm{Decay}\sim$~10$^2$ -- 10$^5$~s observed both in
neutral \citep{Nikolaenko2009,Chekhovich2010} and charged
\citep{Maletinsky2007,Latta2011} III-V quantum dots at high
magnetic field. The long $\tau_\textrm{Decay}$ in III-V QDs are
due to the strain-induced quadrupolar effects which inhibit spin
diffusion via nuclear dipolar flip-flops. Although quadrupolar
effects are absent for the spin $I=1/2$ nuclei in CdTe/ZnTe dots,
the observed $\tau_\textrm{Decay}$ is too short to be ascribed to
spin diffusion alone. The most likely cause of fast DNP decay is
the electron-nuclear interaction \citep{Maletinsky2007,Latta2011}.
Although the PL of the studied QDs is dominated by the neutral
exciton state, the effect of the fluctuating charge environment
(nearby QDs and/or charge traps) is evidenced in spectral
wandering of the PL energy (the wandering differs from dot to dot
and is within $\sim$~100~$\mu$eV over the time scales of hours for
the best QDs, see details in Supplementary Note 4). Moreover, it
is possible that the studied dot itself is intermittently occupied
by electrons or holes leading to nuclear spin depolarization
\citep{Chekhovich2010}. As we show below, NMR spectroscopy
corroborates this interpretation.

\begin{figure}
\includegraphics[width=0.8\columnwidth]{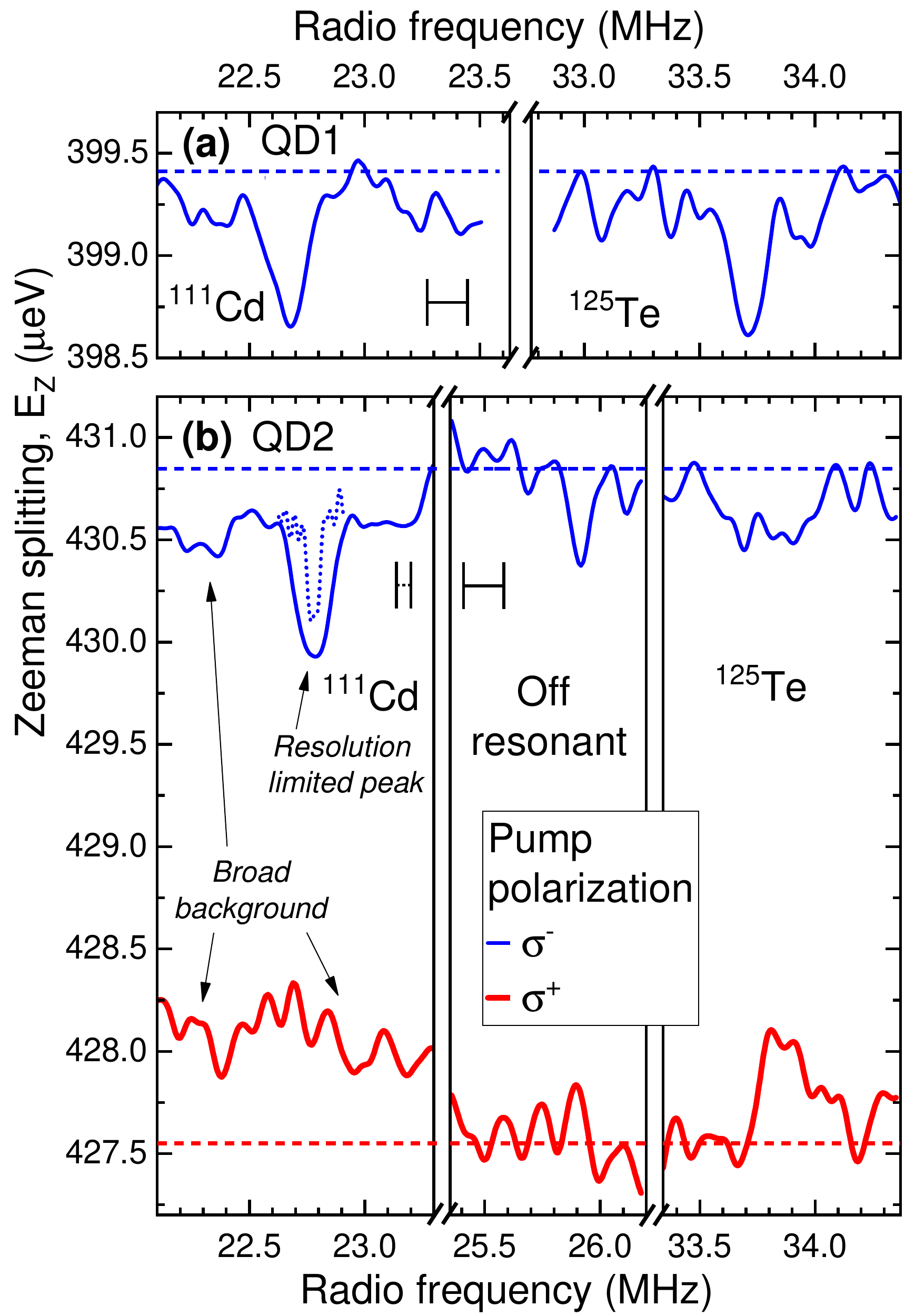}
\caption{Optically detected NMR spectra in CdTe/ZnTe quantum dots
of sample B at $B_{z}\approx$ 2.5 T. Timing diagram of
Fig.~\FigTDiagSpec is used in this measurement with RF pulse
applied during the $T_\textrm{Wait}$. The RF excitation pulse has
a rectangular-band spectrum with a width of 63~kHz (dotted lines)
and 174~kHz (solid lines) which determines spectral resolution
(also shown by the horizontal bars). The RF pulse duration is
$T_\textrm{RF}=$~50 -- 120~ms. In order to increase the signal,
all cadmium NMR spectra are recorded by combining the signals from
$^{111}$Cd and $^{113}$Cd. This is achieved using two RF spectral
bands, where the mean frequency of the second band
$f_\mathrm{^{113}Cd} =
(\gamma_\mathrm{^{113}Cd}/\gamma_\mathrm{^{111}Cd})f_\mathrm{^{111}Cd}$
is the frequency of the main band $f_\mathrm{^{111}Cd}$ scaled by
the ratio of the gyromagnetic values
$\gamma_\mathrm{^{113}Cd}/\gamma_\mathrm{^{111}Cd}=$~9.487/9.069.
(a) Zeeman splitting $E_Z$ in QD1 as a function of radio frequency
measured with a $\sigma^{-}$ pump. Dashed horizontal line shows
$E_Z$ measured without RF. Negative peaks are observed at
$\sim$~22.6~MHz and $\sim$~33.7~MHz corresponding to RF
depolarization of $^{111}$Cd and $^{125}$Te nuclei respectively.
(b) NMR measurements on QD2 with $\sigma^{-}$ (thin blue lines)
and $\sigma^{+}$ (thick red lines) pump. The observed NMR spectra
are a combination of resolution limited peaks and broad
background. Measurements at $\sim$~25.5 -- 26.0~MHz are out of
resonance with all isotopes and demonstrate typical noise
levels.}\label{fig:NMRspec}
\end{figure}

While optical methods can be used to manipulate and detect QD
nuclear spin magnetization along the external field, a complete
control of the magnetization vector requires radio frequency (RF)
magnetic fields. Figure~\ref{fig:NMRspec} shows nuclear magnetic
resonance (NMR) spectra, obtained by depolarizing the nuclei with
RF field of a variable frequency $f_\textrm{RF}$. In order to
balance NMR spectral resolution and signal amplitude, the RF
signal contains multiple frequencies and has a shape of a
rectangular spectral band centered at $f_\textrm{RF}$ (see details
in Supplementary Note 5). Measurement on QD1 conducted at
$B_z\approx$~2.5~T with a $\sigma^{-}$ pump and a resolution of
174~kHz are presented in Fig.~\ref{fig:NMRspec}(a) and show
resolution limited negative peaks at $\sim$~22.6~MHz and
$\sim$~33.7~MHz, which indicate nuclear spin depolarization at the
expected resonance frequencies of $^{111}$Cd and $^{125}$Te. From
the NMR peak amplitude, the total Cd hyperfine shift is
$\sim$~0.8~$\mu$eV. Using the known electron wavefunction density
in CdTe \citep{NAKAMURA1979411}, we find Cd polarization degree
$\gtrsim$~15\%, which is a lower bound since the electron
wavefunction is partly localized in the ZnTe barrier. (NMR has not
been detected on $^{67}$Zn, due to its low abundance and
quadrupolar broadening).

Similar NMR measurements on another individual dot [QD2,
Fig.~\ref{fig:NMRspec}(b)] show a more complex picture. A clear
peak-like structure is observed only for the measurement on Cd
nuclei ($\sim$~22.2 -- 23.3~MHz) with $\sigma^{-}$ pumping.
Measurement with a 174~kHz resolution (solid line) shows a
combination of a resolution limited negative peak
($\sim-0.5$~$\mu$eV amplitude) and a flat background offset of
$\sim-0.3$~$\mu$eV with respect to the Zeeman splitting measured
without RF (horizontal dashed line). As NMR measurements reveal
complex spectra, we have conducted additional measurements in the
$\sim$~25.4 -- 26.2~MHz frequency range [see
Fig.~\ref{fig:NMRspec}(b)] corresponding to RF detuned from all
isotopes. The data reveals only fluctuations with a
$\sim$~0.5~$\mu$eV peak-to-peak amplitude, without any systematic
offset from the reference level measured without RF (dashed
lines). This confirms that the broad ($>$~1~MHz width) background
offsets observed in Cd measurements as well as broad (width
$>$~300~kHz) peaks in Te measurements on QD2 are real NMR signals
and are not related for example to RF-induced sample heating.

The spin-1/2 nuclei are insensitive to electric field gradients,
while the nuclear-nuclear dipolar interactions are limited to few
kHz. This leaves the effective field $B_e$ of the electron spin
(Knight field) as the only source of the broad background in the
NMR spectra. The Knight shift of $^{111}$Cd equals
$\gamma_\mathrm{^{111}Cd}B_e$ and is at least $\sim\pm$~0.5~MHz in
QD2, leading to the estimate $|B_e|\gtrsim$~50~mT. Such a large
$B_e$ can be generated by electrons intermittently occupying the
dot during RF exctiation in the dark. The time-averaged NMR
spectrum of $^{111}$Cd under $\sigma^{-}$ pump [see
Fig.~\ref{fig:NMRspec}(b)] is then explained as a sum of the
narrow peak arising from an empty dot, and a broad offset arising
from the electron-charged state of the dot. Note, that in addition
to the peaks, the broad background is also observed for QD1 in
Fig.~\ref{fig:NMRspec}(a), though to a smaller extent, implying a
smaller fraction of time in an electron-charged state. We further
note that $B_e$ estimated here for CdTe QDs is at least a factor
of $\sim$~5 larger than $|B_e|\sim10$~mT observed in
InGaAs\citep{Lai2006} and InP QDs \citep{Chekhovich2010}, which
agrees with a smaller number of nuclei (with and without spin)
within the volume of the electron in a CdTe QD.

Having established the origin of the broad background we examine
the resolution limited NMR peak. A further measurement of
$^{111}$Cd NMR with a $\sigma^-$ pump and resolution of 63~kHz
[dotted line in Fig.~\ref{fig:NMRspec}(b)] also yields a broad
background offset and a resolution limited peak. However the peak
amplitude is reduced compared to the 174~kHz measurement.
Measurements with even better resolution resulted in amplitude too
small to detect, suggesting that the resolution limited peak
itself consists of a narrow peak (width $\lesssim$~63~kHz) and
broad ($\sim$~100~kHz) wings. The width of the wings suggests
Knight field as the cause, but unlike the broad background, this
smaller broadening of the resolution limited peak is likely to
arise from the Knight field of the electrons occupying nearby
charge traps and/or QDs which are also responsible for spectral
wandering.


In conclusion, we have demonstrated manipulation and probing of
the nuclear spins in individual CdTe quantum dots using optical
and radiofrequency fields. The direct detection of the electron
hyperfine shifts in a pump-probe manner is shown to have a key
role in distinguishing between the real nuclear spin phenomena and
the effects that mimic DNP. Moreover, the direct detection have
enabled exploring arbitrary magnetic fields: at $B_z\gtrsim$~1~T
we achieve fast ($\sim$~1~ms) initialization and long
($\gtrsim$~1~s) persistence of the nuclear spin polarization.
Unlike III-V semiconductor quantum dots where the nuclear spin
bath has a mesoscopic character, II-VI dots offer an attractive
alternative with an inherently smaller number of nuclei
interacting with the electron and a further potential for a
few-spin or spin-free bath via isotope purification. Our results
set the direction for further work required to realize this
potential: Experiments with quasi-resonant and resonant QD optical
excitation can be used to better understand the DNP mechanisms and
achieve highly polarized nuclear spin state. Control of the charge
state of the quantum dot and its environment (e.g. using gated
structures) can overcome inhomogeneous NMR broadening, which in
turn will enable coherent manipulation of the nuclear spins.
Strong electron-nuclear interaction (observed as large Knight
shifts) and the ability to dilute the nuclear spin bath offer in
principle the possibility to address individual nuclear spins with
resonant radiofrequency fields. In this way the II-VI quantum dots
have potential for implementing the hybrid electron-nuclear spin
quantum registers which have been demonstrated in group IV
semiconductors \citep{Dutt2007}, but are not feasible in III-V
dots.

\acknowledgements{This work was supported by the EPSRC Programme
Grant EP/N031776/1. E.A.C. was supported by a University Research
Fellowship from the Royal Society. J.K., W.P., and K.S. were
supported by the Polish National Science Center (grants
DEC-2015/16/T/ST3/00371 and DEC-2015/18/E/ST3/00559).}

\renewcommand{\thesection}{Supplemental Note \arabic{section}}
\setcounter{section}{0}
\renewcommand{\thefigure}{\arabic{figure}}
\renewcommand{\figurename}{Supplemental Figure}
\setcounter{figure}{0}
\renewcommand{\theequation}{\arabic{equation}}
\setcounter{equation}{0}
\renewcommand{\thetable}{\arabic{table}}
\renewcommand{\tablename}{Supplemental Table}
\setcounter{table}{0}

\renewcommand{\citenumfont}[1]{S#1}
\makeatletter
\renewcommand{\@biblabel}[1]{S#1.}
\makeatother

\pagebreak \pagenumbering{arabic}


\section*{Supplemental Material}

The supplemental material contains additional experimental results
supporting the discussion of the main text.



\section{\label{sec:TEM} C\lowercase{d}T\lowercase{e}/Z\lowercase{n}T\lowercase{e} sample structures and trasmission electron microscopy}

We study two CdTe/ZnTe samples grown by molecular beam epitaxy.
The samples were grown on GaAs:Si or GaAs:Zn (100) substrates. The
growth of a ZnTe buffer layer (1~$\mu$m or 1.3~$\mu$m thick), was
followed by atomic layer epitaxy growth of the CdTe layer. Due to
the small energy of dislocation formation in CdTe/ZnTe
\citep{SupTinjod2003}, special measures were taken to achieve
dislocation-free QD formation. Two approaches were used. In sample
A, amorphous Te technique was employed \citep{SupTinjod2003},
where the substrate was strongly cooled (to $\sim$~100~$^\circ$C)
whilst Te was deposited onto the dot layer. This lead to a
decrease of the surface energy and the thin CdTe film layer
subsequently formed the quantum dots \citep{SupJKobakJCG2013}. In
sample B, a 2 nm CdTe quantum well (QW) layer was first deposited
onto the ZnTe buffer. The deposition of amorphous Te was avoided
by reducing the time for which the substrate was cooled under Te
flux. In this way, QD formation was induced while preserving the
quantum well (wetting layer), and the dot density was higher than
in sample A. In both structures, quantum dots were capped by a
100~nm ZnTe layer.

\begin{figure}[h]
\includegraphics[viewport=1 1 562 370, clip=true, width=0.99\linewidth]{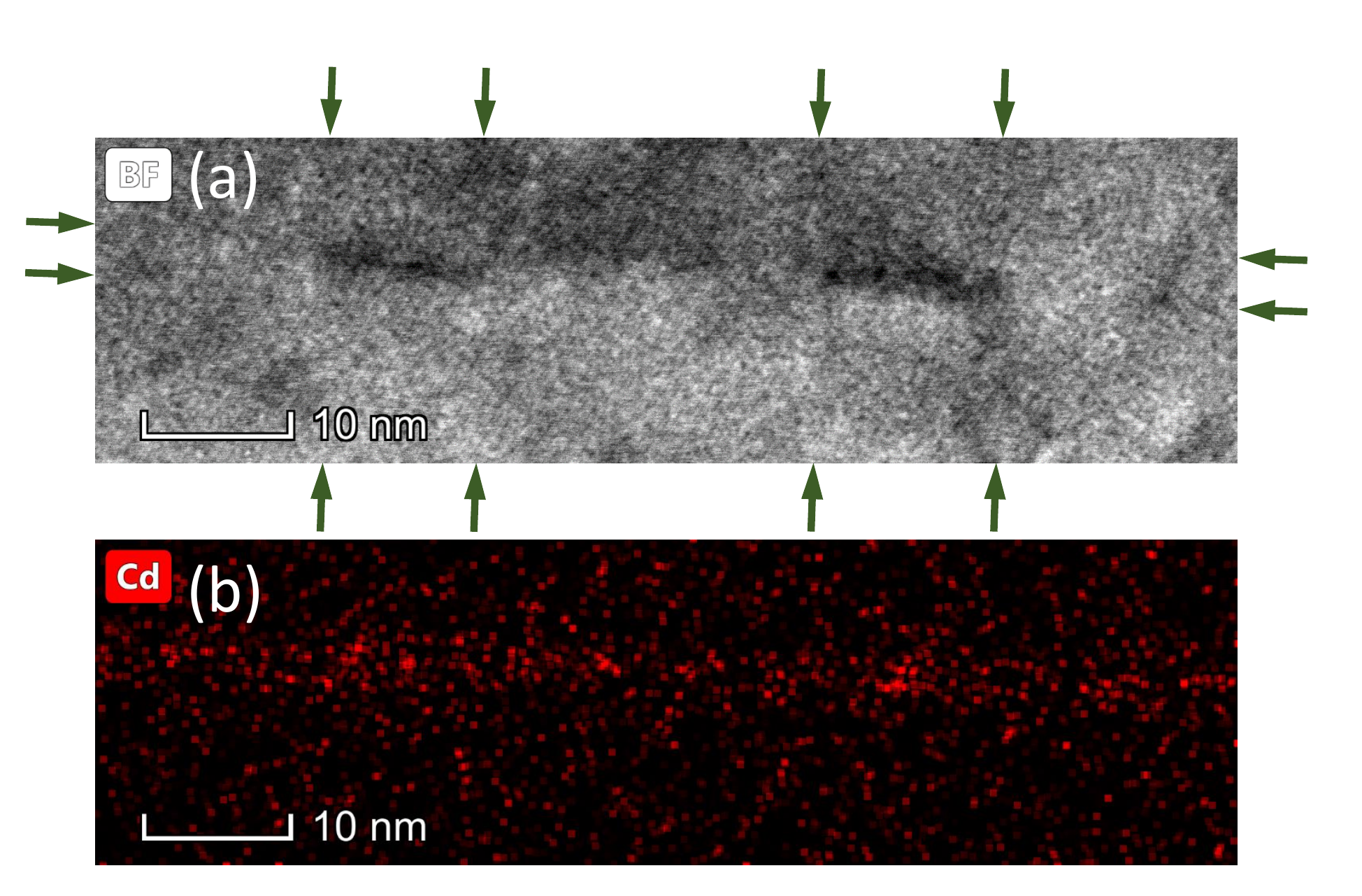}
\caption{\label{Fig:FigSTEM} (a) Representative transmission
electron microscopy (TEM) image taken on sample B under bright
field conditions. The arrows mark the approximate boundary lines
of the two areas attributed to CdTe quantum dots. (b)
Energy-dispersive x-ray (EDX) image of the same sample area as in
(a) showing distribution of cadmium atoms.}
\end{figure}

The structures were examined using transmission electron
microscopy (TEM). A very small amount of cadmium was found in
sample A, complicating quantitative analysis. This agrees with
observation of low QD density in PL experiments on sample A. In
case of sample B, Supplemental Fig.~\ref{Fig:FigSTEM}(a) shows a
representative TEM image taken under bright field conditions. A
clear diffraction contrast is observed as darker areas.
Supplemental Fig.~\ref{Fig:FigSTEM}(b) shows energy-dispersive
x-ray (EDX) image of the same sample as in (a), which reveals that
the darker areas originate from an increased cadmium content. An
approximately horizontal ($\sim$~2$^\circ$ tilt) cadmium-rich
layer is observed in the EDX image and is attributed to the CdTe
quantum well. The EDX image also reveals the inhomogeneity of Cd
within the QW layer. Since the bright field image has better
signal to noise ratio than EDX, we use the former to examine the
scales of fluctuations of cadmium content. The arrows in
Supplemental Fig.~\ref{Fig:FigSTEM}(a) mark the approximate
boundary lines enclosing the two dark areas corresponding to
increased Cd content. We attribute these cadmium-rich areas to QDs
observed in PL experiments, and from the image we find that the
dots have approximately cylindrical shape with a diameter of
$\sim$~10~nm and a height of $\sim$~2.5~nm. Using the lattice
constant of CdTe $a_0=$0.648~nm, and taking into account that
there are 8 atoms per cell, we estimate that a typical quantum dot
contains $\sim$~5000 atoms of all isotopes (with and without
nuclear magnetic moments).

\section{\label{sec:gFactors}Derivation of electron and hole \lowercase{g}-factors}

\begin{figure}[h]
\includegraphics[width=0.99\linewidth]{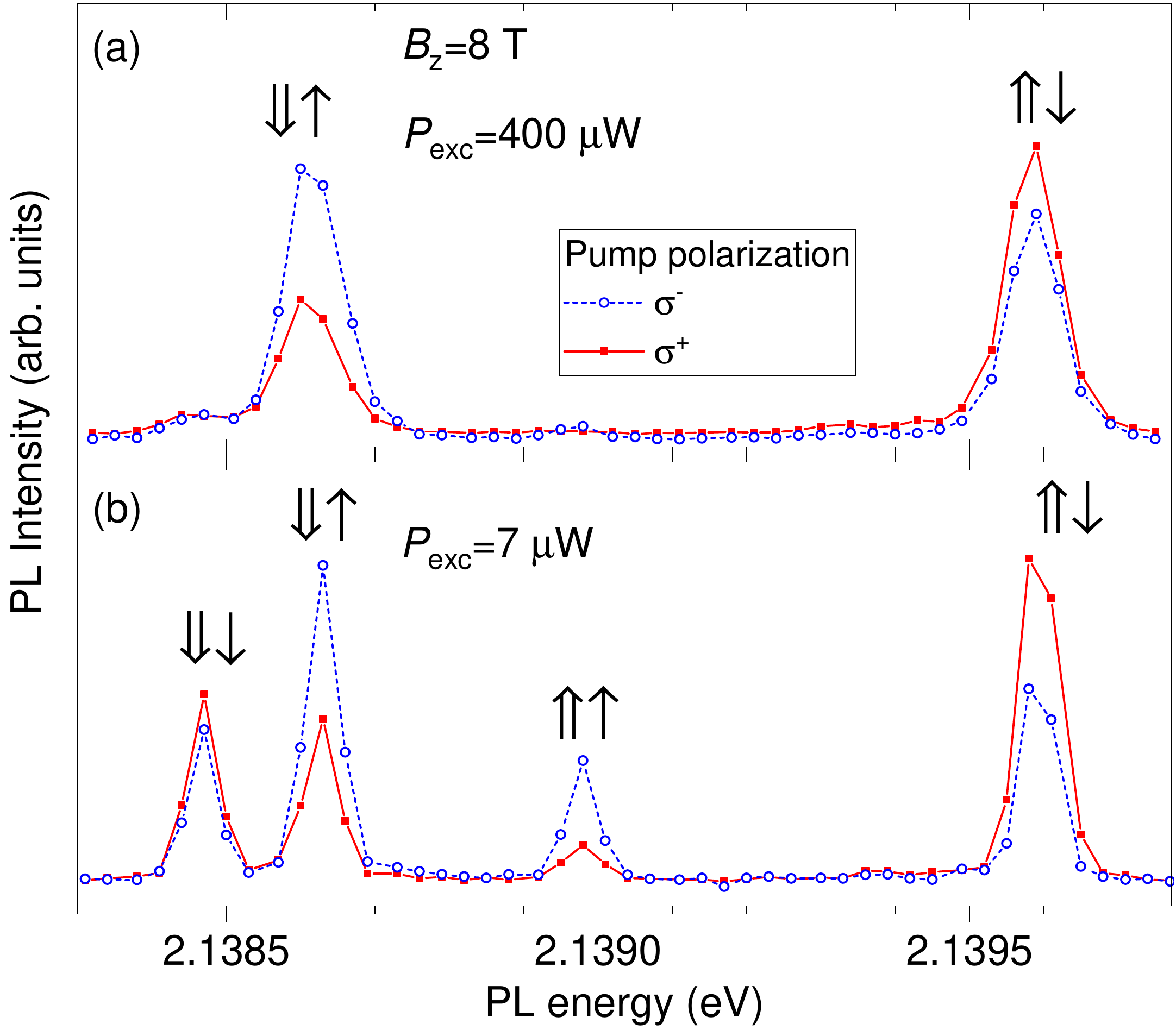}
\caption{\label{Fig:SFiggFactors} Photoluminescence spectra of a
quantum dot in sample A at $B_z$=8~T measured under $\sigma^-$
(dashed lines) and $\sigma^+$ (solid lines) excitation at (a) high
power $P_\textrm{exc}=400$~$\mu$W, and (b) low power
$P_\textrm{exc}=7$~$\mu$W.}
\end{figure}

Supplemental Fig.~\ref{Fig:SFiggFactors} shows photoluminescence
spectra of a single quantum dot in sample A measured in high
magnetic field $B_z=$~8~T. The spectra were recorded under
$\sigma^-$ (dashed lines) and $\sigma^+$ (solid lines) excitation
at two different powers of $P_\textrm{exc}=400$~$\mu$W (a) and
$P_\textrm{exc}=7$~$\mu$W (b). The high power spectrum consist of
a Zeeman doublet corresponding to the bright exciton states
$\Uparrow\downarrow$ and $\Downarrow\uparrow$, where $\Uparrow$
($\Downarrow$) denotes heavy hole states with momentum $j_z=+3/2$
($-3/2$), while $\uparrow$ ($\downarrow$) denotes electron states
with spin $s_z=+1/2$ ($-1/2$). At low power two additional lines
appear -- these correspond to recombination of the nominally dark
states $\Uparrow\uparrow$ and $\Downarrow\downarrow$ made visible
by mixing with the bright states in a low symmetry potential of
the quantum dot \citep{SupPhysRevB.86.241305,SupDNPInP}. From the
low power spectrum we observe that each circularly polarized
excitation enhances one bright and one dark transition. For
example, the $\sigma^-$ excitation enhances the low energy bright
and the high energy dark transition, and since hole spin is
usually lost during relaxation, we can deduce that these two
states have the same electron spin projection. This observation
allows the spectral lines to be assigned to the exciton states.
There are two possible options: one is shown in Supplemental
Fig.~\ref{Fig:SFiggFactors}(b), and the other one has all electron
and hole spin projections reversed. As we now show the correct
assignment can be obtained from electron and hole $g$-factor
calculations.

The energies of the bright excitons $E_{b}$ and dark excitons
$E_{d}$ can be written as follows
\citep{SupBayer,SupBayerDark,SupDNPInP}:
\begin{eqnarray}
\label{eq:Zeeman} E_{b}=E_{0}+\frac{\delta_{0}}{2}\pm\frac{1}{2}\sqrt{\delta^2_{b}+\mu^2_{B}(g_{h}-g_{e})^2B^2_{z}},\nonumber\\
E_{d}=E_{0}-\frac{\delta_{0}}{2}\pm\frac{1}{2}\mu_{B}(g_{h}+g_{e})B_{z},
\end{eqnarray}
where $\mu_{B}$ is Bohr magneton, $E_0$ -- QD band-gap energy,
$g_e$($g_h$) -- electron (hole) g-factor, $\delta_0$ is the
splitting between dark and bright exciton doublets, $\delta_b$ is
the bright doublet fine structure splitting, and we have neglected
the the overall diamagnetic shift and a small dark exciton fine
structure splitting. For the dot shown in Supplemental
Fig.~\ref{Fig:SFiggFactors}, we find $\delta_b\approx60$~$\mu$eV
from PL spectra in low magnetic field. Using this value and the
energies of the four exciton transitions, we calculated the
$g$-factors $g_e\approx\mp0.49$ and $g_h\approx\pm1.59$, where the
two sign combinations correspond to the two possible assignments
of the electron and hole spin states in the PL spectrum. The bulk
electron $g$-factors are negative both in CdTe ($g_e\approx-1.59$,
Ref.~\citep{SupNAKAMURA1979411}) and ZnTe ($g_e\approx-0.6$, Ref.
~\citep{SupSCHMIEDER1983357}), suggesting that it should be
negative in CdTe/ZnTe quantum dots, as was observed
previously~\citep{SupPhysRevB.76.045331}. Thus we conclude that
$g_e\approx-0.49$ and $g_h\approx+1.59$ and the correct assignment
of the spin states is the one shown in Supplemental
Fig.~\ref{Fig:SFiggFactors}(b).

Measurements on several individual QDs in sample A have revealed
very similar values of $g_e$ and $g_h$. Dark exciton emission was
not observed in sample B, but the Zeeman splitting of the bright
states determined by $g_h-g_e$ is found to be very similar to
sample A, suggesting the individual $g_e$ and $g_h$ values are
also similar. Recombination of a $\Uparrow\downarrow$ state
results in emission of a circularly polarized photon with a $+1$
momentum -- throughout this work we label this polarization as
$\sigma^+$. Conversely, excitation in $\sigma^-$ polarization
results in enhanced emission from the $\Downarrow\uparrow$ state.

\section{\label{sec:DNPSign}Derivation of the sign of the nuclear spin polarization}

The Hamiltonian of the hyperfine interaction between the electron
with spin $\mathbf{\hat{s}}$ and a nucleus with spin
$\mathbf{\hat{I}}$ is $\hat{H}_{hf}=A(\mathbf{\hat{s}} \cdot
\mathbf{\hat{I}})$. The hyperfine constant is $A =
(2\mu_0/3)\hbar\gamma_\textrm{N}g_\textrm{e}\mu_\textrm{b}|\psi(0)|^2$
(Ref.~\citep{Supslichter1996}) where $g_\textrm{e}\approx2$ is the
free electron g-factor, $\mu_0$ is magnetic constant,
$\mu_\textrm{b}$ is Bohr magneton, $\hbar$ is Planck constant, and
$|\psi(0)|^2$ is the electron wavefunction density at the site of
a nucleus with gyromagnetic ratio $\gamma_\textrm{N}$. It can be
seen that the sign of $A$ is determined by the sign of
$\gamma_\textrm{N}$. The underlying mechanism of dynamic nuclear
polarization in quantum dots is the spin conserving
electron-nuclear flip-flops made possible by the hyperfine
interaction\citep{SupChekhovich2017}. If electrons with positive
(negative) $z$ spin projection are repeatedly injected into the
dot via optical excitation with circularly polarized light, the
resulting net nuclear spin polarization $\langle I_z\rangle$ is
also positive (negative). Generation of spin polarized electrons
is evidenced for example in PL spectra of Fig.~1(c) and
Supplemental Fig.~\ref{Fig:SFiggFactors} where $\sigma^{+}$
($\sigma^{-}$) optical excitation preferentially enhances the
population and PL intensity of a high- (low-) energy bright
exciton Zeeman state. The spin-conserving nature of the flip-flop
process implies that the signs of the non-equilibrium electron and
nuclear spin polarizations are the same, so that the scalar
product $(\mathbf{\hat{s}} \cdot \mathbf{\hat{I}})$ is always
positive. Thus when nuclear polarization $\langle I_z\rangle$
back-acts on the electron spin via hyperfine interaction, the sign
of the energy shift (Overhauser shift) of the corresponding
exciton state depends only on the sign of $A$. For example, in
III-V semiconductors all nuclei have positive $\gamma_\textrm{N}$
and $A>0$. As a result the exciton state that is populated
preferentially by the circularly polarized light shifts to higher
energy -- this statement is true both for the exciton with
electron spin $s_z=+1/2$ and the exciton with $s_z=-1/2$ as
observed e.g. in GaAs quantum dots \citep{SupPhysRevB.97.235311}.
In the studied CdTe/ZnTe dots all of the Cd and Te isotopes (which
are the most abundant) have $\gamma_\textrm{N}<0$ and hence $A<0$.
Thus the exciton state, whose PL intensity is enhanced under cw
circularly polarized excitation is expected to shift to lower
energy due to the resulting nuclear spin polarization.

In both structures studied here $\sigma^+$ ($\sigma^-$) excitation
enhances PL intensity of the high (low) energy exciton state, as
observed in Fig.~1(c) of the main text for a QD in sample A. The
Overhauser shifts can be probed in pump-probe measurements as
discussed in the main text. For sample B we find that nuclear spin
polarization induced by $\sigma^+$ ($\sigma^-$) excitation
decreases (increases) exciton Zeeman splitting $E_Z$ measured in
pump-probe, i.e. the exciton state whose population is enhanced by
circularly polarized excitation shifts to lower energy. This
agrees with analysis above and confirms that nuclear spin
polarization in sample B is produced by spin-conserving
electron-nuclear flip-flops. By contrast, cw experiments on sample
A show that Zeeman splitting under $\sigma^+$ pumping can both
increase and decrease depending on the individual quantum dot.
This further confirms that the spectral shifts observed in sample
A are not related to nuclear spin effects.

\section{\label{sec:SpecWand}Spectral wandering in photoluminescence of the C\lowercase{d}T\lowercase{e}/Z\lowercase{n}T\lowercase{e} quantum dots}

\begin{figure}[h]
\includegraphics[width=0.99\linewidth]{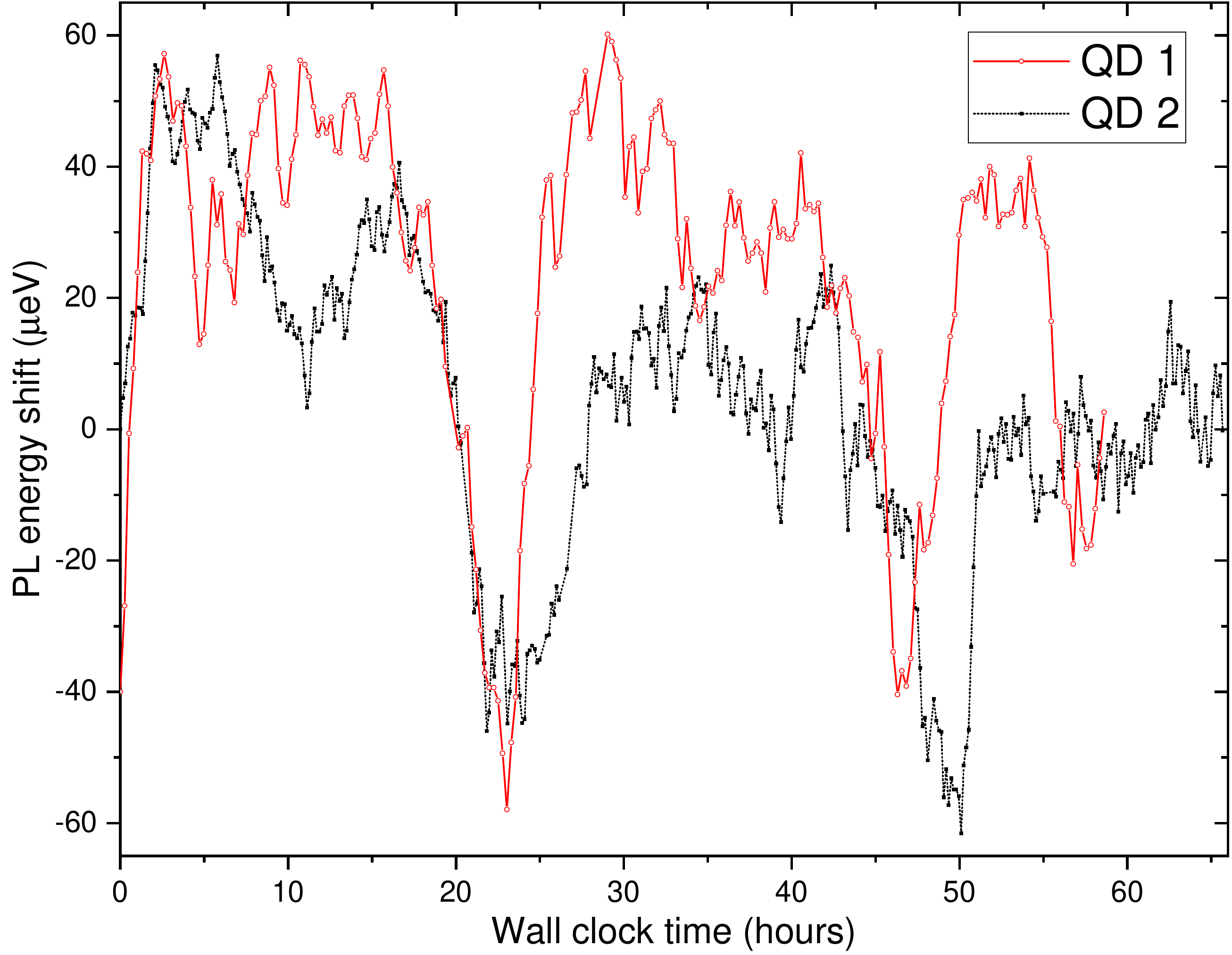}
\caption{\label{Fig:SFigSpecWand} Variation of the
photoluminescence energy of the ground state neutral exciton as a
function of the wall clock time under optical excitation. Data
shown for two separate experiments on two different quantum dots.}
\end{figure}

Supplemental Fig.~\ref{Fig:SFigSpecWand} shows variation of
photoluminescence energies over the experiment duration (NMR
spectra measurements). The results are shown for two different
quantum dots (QD1 and QD2) from sample B, and it can be seen that
spectral wandering is within $\sim$100~$\mu$eV over the time
scales of hours. Spectral wandering\citep{Supdoi:10.1063/1.117565}
arises from the changes in the charge
environment\citep{SupPhysRevLett.108.107401} induced by continuous
optical excitation, and reduces the accuracy with which the
spectral splitting can be deduced from the photoluminescence
signal. In the studied structures spectral wandering was found to
differ significantly from dot to dot. Thus for the detailed
studies of the nuclear spin phenomena, where small changes in
spectral splitting need to be measured accurately, we have
selected quantum dots with minimal wandering, such as those shown
in Supplemental Fig.~\ref{Fig:SFigSpecWand}.

\section{\label{sec:NMR}Details of NMR techniques}

Optically detected nuclear magnetic resonance (NMR) spectra are
measured using a pump-probe protocol consisting of the following
steps: (i) nuclear spins are first polarized using circularly
polarized light, (ii) a radio frequency (RF) pulse is applied
without optical excitation to depolarize the nuclei selectively,
(iii) nuclear spin polarization is finally detected by measuring
QD PL under a short probe laser pulse. This cycle is typically
repeated several hundred times for each radio frequency in order
to accumulate probe PL signal sufficient for accurate derivation
of the PL spectral splitting.

We use the ``saturation NMR'' measurement, where the frequency of
a weak RF field is scanned to record the spectrum. The main
difference from conventional saturation NMR is that a harmonic RF
field of frequency $f_\textrm{RF}$ is replaced by a RF field with
a rectangular spectral band centered at $f_\textrm{RF}$. This
spectral band is approximated by a ``frequency comb'' which is a
sum of equally spaced harmonic modes. The mode spacing in our
experiments is kept at $f_\textrm{ms}=$~125~Hz. The total width of
the band formed by the comb $\textrm{w}_\textrm{exc}$ is chosen to
be between few kHz and few hundred kHz depending on the
measurement. This approach has been used previously for NMR
spectroscopy of III-V semiconductor quantum dots
\citep{SupChekhovich2012}. The advantage of the rectangular band
RF is that $\textrm{w}_\textrm{exc}$ can be varied to balance NMR
signal amplitude and spectral resolution. For example, with a
larger $\textrm{w}_\textrm{exc}$ a larger fraction of nuclei is
depolarized at each $f_\textrm{RF}$, resulting in a larger change
in spectral splitting and hence improved NMR signal to noise
ratio. At the same time, all NMR spectral features narrower than
$\textrm{w}_\textrm{exc}$ are averaged out, so that larger
$\textrm{w}_\textrm{exc}$ limits the resolution. The duration of
the RF pulse $T_\textrm{RF}$ in NMR spectral measurements is
chosen based on an additional experiment with large
$\textrm{w}_\textrm{exc}$ covering the entire resonance and
variable $T_\textrm{RF}$: the exponential depolarization time of
the nuclear spins $\tau$ is derived from such a calibration
measurement and $T_\textrm{RF}$ is set to $\sim$~5$\tau$ for NMR
spectroscopy.

Cadmium has two stable spin-1/2 isotopes: $^{111}$Cd with
gyromagnetic ratio
$\gamma_\mathrm{^{111}Cd}/(2\pi)\approx-9.06915$~MHz/T and
$^{113}$Cd with
$\gamma_\mathrm{^{113}Cd}/(2\pi)\approx-9.48709$~MHz/T
(Ref.\citep{SupHarris2002}). Within the volume of a quantum dot
the two types of Cd are distributed randomly so that on average,
both isotopes experience the same statistical distributions of the
chemical shifts and Knight fields. As a result, the NMR spectral
lineshape of $^{113}$Cd is approximately the NMR spectrum of
$^{111}$Cd but with all frequencies multiplied by
$\gamma_\mathrm{^{113}Cd}/\gamma_\mathrm{^{111}Cd}$. Here we use
this property to increase the magnitude of the NMR signal by
measuring the response of $^{111}$Cd and $^{113}$Cd
simultaneously. In Cd NMR experiments the spectrum of the RF
excitation consists of two rectangular bands (frequency combs) of
the same intensity centered at frequencies $f_\textrm{RF}$ and
$(\gamma_\mathrm{^{113}Cd}/\gamma_\mathrm{^{111}Cd})f_\textrm{RF}$.
The value of $f_\textrm{RF}$ is stepped in the experiment and is
plotted on the horizontal axis of Fig.~4 of the main text. The
resulting Cd NMR spectra correspond to the $^{111}$Cd NMR
lineshape which is amplified by adding a $^{113}$Cd lineshape with
properly rescaled frequencies.

\end{document}